\documentclass[12pt]{article}
\title{
\vspace{-3mm}
\rightline{\small IFUP-TH 2002/29}
\vspace{8mm}
\bf Exploring quark-gluon plasma on the loop space}
\author{
Dmitri Antonov \thanks{
E-mail: {\tt antonov@df.unipi.it}}
\thanks{Permanent address:
ITEP, B. Cheremushkinskaya 25, RU-117 218 Moscow, Russia.}\\
{\it INFN-Sezione di Pisa, Universit\'a degli studi di Pisa,
Dipartimento di Fisica,}\\
{\it Via Buonarroti, 2 - Ed. B -
I-56127 Pisa, Italy}}

\date{}
\begin{document}


\maketitle
\vspace{1mm}
\centerline{\bf {Abstract}}
\vspace{3mm}
\noindent
Langevin equation describing soft modes in the quark-gluon plasma is reformulated on the
loop space. The Cauchy problem for the resulting loop equation is solved for the
case when the nonvanishing components of the gauge potential correspond to the Cartan generators
of the $SU(N)$-group and are proportional to a constant unit vector in the Cartan subalgebra.
The regularized form of the loop equation with an arbitrary gauge potential
is found, and perturbation theory in powers of the
't~Hooft coupling is discussed.

\vspace{5mm}

\vspace{5mm}

\vspace{10mm}

\section{Introduction}
Loop equations in QCD were derived about 20 years ago~\cite{le} (for a review see ref.~\cite{rev}),
and nowadays there exist numerous applications of the loop-space techniques to other theories in $D>2$. Among those, apart from
various proposals of string theories obeying loop equations~\cite{str, dgo}, it is worth mentioning
applications to supersymmetric gauge theories~\cite{dgo, susy}, gravity~\cite{gravity}, and
turbulence~\cite{tur1, tur2}. In the present letter, we shall derive and analyse a novel loop equation
describing the dynamics of soft gauge fields in quark-gluon plasma. It is based on the following Langevin-type
equation~\cite{bod}

\begin{equation}
\label{1}
\gamma\partial_t A_i=-D_jF_{ji}+g\zeta_i,
\end{equation}
where the bilocal correlator of
Gaussian noises reads

\begin{equation}
\label{cO}
\left<\zeta_i^a(x)\zeta_j^b(y)\right>=2T\gamma\delta_{ij}\delta^{ab}\delta(x-y).
\end{equation}
In these equations, $i,j=1,2,3$, $a=1,\ldots,N^2-1$, $\gamma$ is the so-called color conductivity,
whose value in the pure $SU(N)$-theory under study is~\cite{sg}

\begin{equation}
\label{gam}
\gamma=\frac{4\pi T}{9\ln g^{-1}},
\end{equation}
$T$ is the temperature,
$A_i=igA_i^at^a$ is an anti-Hermitean matrix with
${\rm tr}{\,}t^at^b=\delta^{ab}$, $D_i=\partial_i+A_i$, and $F_{ij}=\partial_iA_j-\partial_jA_i+[A_i,A_j]$
(here we have adapted the notations of refs.~\cite{le, rev}).

Let us briefly quote the qualitative derivation
of eqs.~(\ref{1})-(\ref{cO}) presented in ref.~\cite{asy}. It is based on the idea that soft modes are effectively
classical and therefore can
be described by the non-Abelian analogue of the Maxwell equation
$\vec D\times\vec B=D_t\vec E+\vec J_{\rm hard}$. Here, $\vec B=\vec D\times\vec A$ and
all the covariant derivatives involve only soft degrees of freedom, while the color current $\vec J_{\rm hard}$
describes hard degrees of freedom. Owing to the fact that plasmas are conductors, this current can be expressed in terms
of $\vec E$ as $\vec J_{\rm hard}=\gamma\vec E$. Assuming that the characteristic temporal scale of soft modes is much
larger than their spatial scale (that will be discussed later) and adapting the gauge $A_0=0$, we arrive at the
following equation: $\gamma\partial_t\vec A=-\vec D\times\vec B$. This equation is dissipative, and the dissipation
is a consequence of interaction of the soft modes with the hard ones. However, hard modes not only acquire the energy from
the soft modes, but also serve as a source of thermal noise $\vec\zeta$. The condition necessary to reach the
equilibrium distribution of soft modes is the balance between their excitation by the thermal noise and dissipative decay.
For a general Langevin equation of the type $\gamma\partial_t\vec q=-\nabla_{\vec q}V(\vec q{\,})+\vec\zeta$,
one can show~\cite{zj} that the correct equilibrium distribution ${\rm e}^{-V/T}$ is reproduced by the Gaussian noise,
whose correlation functions read

$$\left<\zeta_i(t)\zeta_j(t')\right>=2T\gamma\delta_{ij}\delta(t-t'),~~
\left<\prod\limits_{a=1}^{2k+1}\zeta_{i_a}(t_a)\right>=0,~~
\left<\prod\limits_{a=1}^{2k}\zeta_{i_a}(t_a)\right>=\sum\limits_{{\rm possible~ pair}\atop {\rm combinations}}^{}
\prod\limits_{\rm pairs}^{}\left<\zeta_{i_a}\left(t_a\right)\zeta_{i_b}\left(t_b\right)\right>.$$
These are the qualitative considerations that lead to the effective theory for soft modes, described by eqs.~(\ref{1})-(\ref{cO}).

Let us also discuss the characteristic spatial and temporal scales of soft fields. Since we are interested in {\it soft}
(or nonperturbative) fields, both terms in the covariant derivative associated with such fields, should be equally important.
This means the inequality $A\ge {\cal O}(1/gR)$, where $A$ is the amplitude of a typical nonperturbative
fluctuation of the spatial size $R$. The energy of such a fluctuation is thus $E\sim RA^2\ge{\cal O}(1/g^2R)$.
For the probability of fluctuation not to be suppressed as ${\rm e}^{-E/T}\sim{\rm e}^{-\frac{1}{g^2RT}}$, one
needs $R\ge{\cal O}(1/g^2T)$. Therefore, the characteristic spatial scale of soft fields is

\begin{equation}
\label{spa}
R\sim\frac{1}{g^2T}.
\end{equation}
Next, from eq.~(\ref{1}), we have the following estimate involving the temporal scale $\tau$ of soft fields: $\gamma\tau^{-1}A\sim
R^{-2}A$, so that $\tau\sim\gamma R^2\sim\gamma/\left(g^4T^2\right)$. By virtue of eq.~(\ref{gam}),
we finally have

\begin{equation}
\label{tem}
\tau\sim\frac{1}{g^4T\ln g^{-1}}.
\end{equation}
In particular, this yields the estimates

$$
\frac{\tau^{-1}}{\gamma}\sim g^4\left(\ln g^{-1}\right)^2\ll1,~~
\frac{\tau^{-1}}{R^{-1}}\sim g^2\ln g^{-1}\ll1,$$
where the inequalities are obviously valid for small enough $g$. These inequalities justify
the omission of the term $D_t\vec E$ in the equation $\vec D\times\vec B=D_t\vec E+\gamma\vec E$ above.

The letter is organized as follows. In the next Section, we shall present the derivation of the large-$N$ loop equation in
quark-gluon plasma. In Section~3, the Cauchy problem for this equation will be solved at a certain class of fields.
In Section~4, the regularized version of the equation will be derived, and the perturbative expansion will be briefly
discussed. Finally, the main results of the letter will be presented in Summary.

\section{Derivation of the loop equation in quark-gluon plasma}

Similarly to the theory of turbulence~\cite{tur1}, the Wilson loop,

$$W\equiv W[C,t]=\frac{1}{N}\left<{\rm tr}{\,}{\cal P}{\,}\exp
\left(\oint\limits_{C}^{} dx_i A_i({\bf x},t)\right)\right>,$$
is a functional of the 3D contour $C$ and a function of time. It it worth emphasizing that the average here is implied in the
sense of the average over the Gaussian noise,

$$\left<{\cal O}\right>=\frac{\int {\cal D}\zeta_i^a{\cal O}\exp\left[-\frac{1}{4T\gamma}\int d^3x\int\limits_{0}^{\infty}dt
\zeta_i^{a{\,}2}\right]}{\int {\cal D}\zeta_i^a\exp\left[-\frac{1}{4T\gamma}\int d^3x\int\limits_{0}^{\infty}dt
\zeta_i^{a{\,}2}\right]},$$
rather than over the Yang-Mills action.

Differentiating $W$ with respect to time
and using eq.~(\ref{1}) we get

\begin{equation}
\label{2}
\left(\gamma\partial_t+\Delta\right)W=\frac{g}{N}
\oint\limits_{C}^{} dx_i{\,}{\rm tr}{\,}\left<\zeta_i({\bf x},t)U_{\bf xx}\right>.
\end{equation}
Here, $\Delta\equiv\int_{0}^{1}d\sigma\int_{\sigma-0}^{\sigma+0}d\sigma'\frac{\delta^2}{\delta x_i(\sigma')
\delta x_i(\sigma)}$ is the functional Laplacian~\footnote{Rigorously speaking, this form of the functional Laplacian is
admissible only when it acts onto functionals like $W$, which do not have marked points.} and

\begin{equation}
\label{u}
U_{\bf xy}\equiv{\cal P}{\,}\exp
\left(\int\limits_{C_{\bf xy}}^{} dz_i A_i({\bf z},t)\right)
\end{equation}
is the parallel transporter factor along the
piece $C_{\bf xy}$ of the contour $C$ which starts at the point ${\bf y}$ and ends up at the point ${\bf x}$.
The correlator ${\rm tr}{\,}\left<\zeta_i({\bf x},t)U_{\bf xx}\right>$ on the r.h.s. of eq.~(\ref{2})
can further be rewritten as follows~\footnote{Although we do not expect any confusion between the space and matrix indices, $i,j$,
note that the former are always denoted by the lower case, while the latter are denoted by the upper one.}:

$$
(t^a)^{ij}\left<\zeta_i^a({\bf x},t)U_{\bf xx}^{ji}\right>=
2T\gamma(t^a)^{ij}\left<\frac{\delta}{\delta\zeta_i^a({\bf x},t)}U_{\bf xx}^{ji}\right>=
2T\gamma(t^a)^{ij}\int d^3y\left<\frac{\delta A_j^b({\bf y},t)}{\delta\zeta_i^a({\bf x},t)}
\frac{\delta}{\delta A_j^b({\bf y},t)}U_{\bf xx}^{ji}\right>.$$
Taking into account that~\cite{zj}

\begin{equation}
\label{deriv}
\frac{\delta A_j^b({\bf y},t)}{\delta\zeta_i^a({\bf x},t)}=\frac{g}{2\gamma}
\delta_{ij}\delta^{ab}\delta({\bf x}-{\bf y}),
\end{equation}
we arrive at the expression
$gT(t^a)^{ij}\left<\frac{\delta}{\delta A_i^{a}({\bf x},t)}
U_{\bf xx}^{ji}\right>$. It can be evaluated by virtue of the formulae

$$
\frac{\delta U_{\bf zy}^{ji}}{\delta A_i^a({\bf x},t)}=\int\limits_{C_{\bf zy}}^{}du_jU_{\bf zu}^{jk}(t^b)^{kl}
\frac{\delta A_j^b({\bf u},t)}{\delta A_i^a({\bf x},t)}U_{\bf uy}^{li}=(t^a)^{kl}\int\limits_{C_{\bf zy}}^{}du_i\delta({\bf u}-{\bf x})
U_{\bf zu}^{jk}U_{\bf uy}^{li}$$
and

\begin{equation}
\label{tt}
(t^a)^{ij}(t^a)^{kl}=\delta^{il}\delta^{kj}-\frac{1}{N}\delta^{ij}\delta^{kl}.
\end{equation}
Using the large-$N$ factorization of the resulting two-loop average,
we eventually arrive at the
following large-$N$ loop equation equivalent to the original eq.~(\ref{1}):

\begin{equation}
\label{3}
\left(\gamma\partial_t+\Delta\right)W=\lambda T\oint\limits_{C}^{}dx_i\oint\limits_{C}^{}dy_i\delta({\bf x}-{\bf y})
W[C_{\bf xy},t]W[C_{\bf yx},t].
\end{equation}
Here, $\lambda=g^2N$ is the 't~Hooft coupling constant which remains finite in the large-$N$ limit,
and the contours $C_{\bf xy}$, $C_{\bf yx}$ are always closed due to the presence of the $\delta$-function.
Note that functional form of the r.h.s. of eq.~(\ref{3}) is the same as that of the usual loop equation~\cite{le},
while the l.h.s. is different due to the presence of the temporal derivative.

\section{Solution of the Cauchy problem for a certain class of fields}

The Cauchy problem for eq.~(\ref{3}) with the initial condition $W[C,t=0]=W_0[C]$ can be solved for the case when the
only nonvanishing color components of the gauge potential correspond to the Cartan generators
${\bf H}=\left(H^1,\ldots,H^{N-1}\right)$ of the group $SU(N)$, i.e., $A_i={\bf A}_i{\bf H}$, and moreover ${\bf A}_i={\bf n}\psi_i$. Here,
${\bf n}$ is a constant unit vector in the Cartan subalgebra, and $\psi_i({\bf x}, t)$ is an arbitrary 3D vector-function.
Introducing the notations $\varphi_{\bf xy}=\int_{C_{\bf xy}}^{}dz_i\psi_i({\bf z},t)$,
$\varphi_{\bf yx}=\int_{C_{\bf yx}}^{}dz_i\psi_i({\bf z},t)$ and making use of the obvious formula
${\rm e}^{i\varphi{\bf n}{\bf H}}=\cos\varphi+i{\bf n}{\bf H}\sin\varphi$ (which stems from the orthonormality of roots~\cite{group},
$H_iH_j=\delta_{ij}$), we have in the large-$N$ limit under study

$$W[C_{\bf xy},t]W[C_{\bf yx},t]=\left<{\rm e}^{i\varphi_{\rm xy}}\right>\left<{\rm e}^{i\varphi_{\rm yx}}\right>
\simeq\left<{\rm e}^{i\left(\varphi_{\rm xy}+\varphi_{\rm yx}\right)}\right>=W.$$
Denoting further $V=\ln W$ and using the fact the the functional Laplacian obeys the Leibnitz rule, we can rewrite
eq.~(\ref{3}) as

\begin{equation}
\label{Abel}
\left(\gamma\partial_t+\Delta\right)V\left[\vec x{\,},t\right]=J\left[{\vec x}{\,}\right],
\end{equation}
where $J\left[\vec x{\,}\right]\equiv\lambda T\oint_{C}^{}dx_i\oint_{C}^{}dy_i\delta({\bf x}-{\bf y})$ and from now on we shall use
an alternative notation ${\cal O}[C]\equiv{\cal O}\left[\vec x{\,}\right]$, so that $\vec x\equiv\vec x(\sigma)$, $0\le\sigma\le 1$
is an element of the loop space.

In order to invert the operator standing on the l.h.s. of eq.~(\ref{Abel}) (that would lead to
the solution of the Cauchy problem for this equation) we
shall make use of the method proposed
in ref.~\cite{lapl}. Note that this method has been applied in ref.~\cite{tur2} in order to solve the Cauchy problem for the
loop equation in turbulence, and here we shall use the same strategy.
In ref.~\cite{lapl}, it has been shown that for an arbitrary functional ${\cal O}\left[\vec x{\,}\right]$
defined on the loop space, the following equation holds

\begin{equation}
\label{4}
\Delta^{(G)}\left<{\cal O}\left[\vec x+\sqrt{A}\vec\xi{\,}\right]\right>_{\vec\xi}=
2\frac{d}{dA}\left<{\cal O}\left[\vec x+\sqrt{A}\vec\xi{\,}\right]\right>_{\vec\xi}.
\end{equation}
In this equation, the smeared Laplacian $\Delta^{(G)}$ has the form

\begin{equation}
\label{5}
\Delta^{(G)}=\int\limits_0^1 d\sigma{\,} \textsf{v.p.}\int\limits_0^1 d\sigma^\prime
G\left(\sigma-\sigma^\prime\right)\frac{\delta^2}{\delta x_i\left(
\sigma^\prime\right)
\delta x_i\left(\sigma\right)}+\Delta,
\end{equation}
where $\textsf{v.p.}\int d\sigma'\equiv
\int_{0}^{\sigma-0}d\sigma'+
\int_{\sigma+0}^{1}d\sigma'$,
and the average over loops is defined as follows

\begin{equation}
\label{average}
\left<{\cal O}\left[\vec \xi{\,}\right]\right>_{\vec\xi}=
\frac{\int\limits_{\vec\xi(0)=
\vec\xi(1)}^{}{\cal D}\vec\xi e^{-S}{\cal O}\left[\vec\xi{\,}\right]}
{\int\limits_{\vec
\xi(0)=\vec\xi(1)}^{}{\cal D}\vec\xi e^{-S}}.
\end{equation}
Here, $G\left(\sigma-\sigma^\prime\right)$ is a certain smearing function and

\begin{equation}
\label{6}
S=\frac{1}{2}\int\limits_0^1 d\sigma\int\limits_0^1 d\sigma^
\prime\vec\xi\left(\sigma\right)G^{-1}\left(\sigma-\sigma^\prime
\right)\vec\xi\left(\sigma^
\prime\right)
\end{equation}
with $G^{-1}$ standing for the inverse operator.
Note that the first term on the r.h.s. of eq.~(\ref{5}) is an operator of the second
order (i.e., it does not obey the Leibnitz rule) and is reparametrization noninvariant.

In particular, for

\begin{equation}
\label{7}
G\left(\sigma-\sigma^\prime\right)={\rm e}^{-\frac{\left|\sigma-
\sigma^\prime\right|}
{\varepsilon}},~
\varepsilon\ll 1,
\end{equation}
the action~(\ref{6}) becomes local: $S=\frac{1}{4}\int\limits_0^1
d\sigma\left[\varepsilon\left(\partial_\sigma\vec\xi{\,}\right)^2+\frac{1}{\varepsilon}
{\vec\xi}{\:}^2\right]$. For $G\left
(\sigma-\sigma^\prime\right)$ defined
by eq.~(\ref{7}), the contribution of the first term on the r.h.s. of eq.~(\ref{5}) is of
order $\varepsilon$ for smooth contours, and therefore
when $\varepsilon\to 0$ it vanishes,
$\Delta^{(G)}$ tends to the standard functional Laplacian
$\Delta$, and reparametrization invariance becomes restored. In general, the smearing function
$G(\sigma)$ obeys the relations $G(\sigma\ne 0)\to 0$ at $\varepsilon\to 0$, $G(0)=1$ and
can be obtained from the one given by eq.~(\ref{7}) upon a reparametrization.
Equation~(\ref{4}) is then a consequence of the equation of motion

\begin{equation}
\label{eqm}
\left<\xi_i(\sigma){\cal O}\left[\vec\xi{\,}\right]\right>_{\vec\xi}=
\int\limits_0^1 d\sigma^\prime G\left(\sigma-\sigma^\prime\right)\left<\frac{
\delta{\cal O}\left[\vec\xi{\,}\right]}{\delta\xi_i\left(\sigma^\prime
\right)}\right>_{\vec\xi},
\end{equation}
which follows from eqs.~(\ref{average}) and~(\ref{6}).

Performing a shift of the contour $C$, $\vec x\to\vec x+\sqrt{A}\vec \xi$
in eq.~(\ref{Abel}) and using eq.~(\ref{4}), we arrive at the following
equation:

$$
\left(\gamma\frac{\partial}{\partial t}+2\frac{\partial}{\partial A}\right)
\left<V\left[\vec x+\sqrt{A}\vec\xi, t\right]\right>_{\vec\xi}=\left<J\left[\vec x+\sqrt{A}\vec\xi{\,}
\right]\right>_{\vec\xi}.$$
The solution to the Cauchy problem for this equation obviously reads

$$
\left<V\left[\vec x+\sqrt{A}\vec\xi, t\right]\right>_{\vec\xi}=
\left<V\left[\vec x+\sqrt{A-\frac{2t}{\gamma}}\vec\xi, 0\right]\right>_{\vec\xi}+\frac{1}{\gamma}
\int\limits_{0}^{t}d\tau\left<J\left[\vec x+\sqrt{A-\frac{2}{\gamma}(t-\tau)}\vec\xi{\,}\right]\right>_{\vec\xi}.$$
Taking further into account that for nonperturbative fields under study,
the minimal area associated with the contour $C$, $S_{\rm min}$, obeys the inequality
$S_{\rm min}\gg R^2$ and recalling eqs.~(\ref{gam})-(\ref{tem}), we obtain the following estimate:

$$\left|\frac{t}{\gamma}\Delta\left<V\left[\vec x+\sqrt{A}\vec\xi, 0\right]\right>_{\vec\xi}\right|
\sim\frac{\tau}{\gamma S_{\rm min}}\ll\frac{T}{\gamma\ln g^{-1}}=\frac{9}{4\pi}\simeq 0.72.$$
Using this smallness and eq.~(\ref{eqm}), it is straightforward to get the following formula:

$$
\left<V\left[\vec x+\sqrt{A-\frac{2t}{\gamma}}\vec\xi, 0\right]\right>_{\vec\xi}=\left[1-\frac{t}{\gamma}
\Delta^{(G)}+{\cal O}\left(\left(\frac{\tau}{\gamma S_{\rm min}}\right)^2\right)\right]
\left<V\left[\vec x+\sqrt{A}\vec\xi, 0\right]\right>_{\vec\xi}.
$$
Obviously, an analogous expansion can be derived for
$\left<J\left[\vec x+\sqrt{A-\frac{2}{\gamma}(t-\tau)}\vec\xi{\,}\right]\right>_{\vec\xi}{\,}$, after which one can send the parameter
$\varepsilon$ to zero and take into account that $\Delta J\left[\vec x{\,}\right]=0$~\footnote{This fact makes the expansion for
the current $J\left[\vec x{\,}\right]$ exact at the zeroth order
of the parameter $\frac{t}{\gamma S_{\rm min}}$.}. Setting finally the parameter $A$ to be equal zero (that removes the
$\vec\xi$-dependence completely) and passing from $V$
back to $W$, we arrive at the following solution to the Cauchy problem for eq.~(\ref{Abel}):

\begin{equation}
\label{sOl}
W\left[\vec x,t\right]=W_0\left[\vec x{\,}\right]\exp\left[\frac{t}{\gamma}\left(J\left[\vec x{\,}\right]-
\frac{\Delta W_0\left[{\vec x}{\,}\right]}{W_0\left[{\vec x}{\,}\right]}\right)+
{\cal O}\left(\left(\frac{\tau}{\gamma S_{\rm min}}\right)^2\right)\right].
\end{equation}
In particular, by virtue of the formula~\cite{le, rev}

\begin{equation}
\label{D0}
\Delta\oint\limits_{C}^{}dx_i\oint\limits_{C}^{}dy_iD_0({\bf x}-{\bf y})=
-2\oint\limits_{C}^{}dx_i\oint\limits_{C}^{}dy_i\delta({\bf x}-{\bf y}),
\end{equation}
where $D_0({\bf x})=(4\pi |{\bf x}|)^{-1}$ is the
massless propagator,
it is straightforward to see that at $W_0\left[\vec x{\,}\right]=\exp\left(-\frac{\lambda T}{2}
\oint_C^{}dx_i\oint_C^{}dy_iD_0({\bf x}-{\bf y})\right)$,
$W\left[\vec x,t\right]=W_0\left[\vec x{\,}\right]\exp\left[
{\cal O}\left(\left(\frac{\tau}{\gamma S_{\rm min}}\right)^2\right)\right]$. It is also worth noting once more
that the obtained solution~(\ref{sOl}) is valid only at $t\le{\cal O}(\tau)$, rather than at an arbitrarily large $t$.

\section{Regularized loop equation and perturbative analysis}
In order to regularize eq.~(\ref{3}), one should use the regularized version of eq.~(\ref{1}). It is analogous to the one proposed for
the method of stochastic quantization in ref.~\cite{bhst} and reads

$$\gamma\partial_t A_i^a({\bf x},t)=-D_j^{ab}F_{ji}^b({\bf x},t)+g\int d^3yR(D^2)_{\bf xy}^{ab}\zeta_i^b({\bf y},t).$$
Here,
$R(D^2)_{\bf xy}^{ab}=\left({\rm e}^{\frac{D^2}{4\Lambda^2}}\right)^{ab}\delta({\bf x}-{\bf y})$
is the regularizing function with $\Lambda$ standing for the UV cutoff. An analogous method of regularization of the usual
loop equation has been used in ref.~\cite{mh}, where the equality of $-\Delta$ to the field-space Laplacian
(implied in the weak sense) has been employed. Here, we shall perform the regularization directly, without
the use of correspondence between the two Laplacians. One has

$$
R(D^2)_{\bf xy}^{ab}=\int\limits_{{{\bf r}\left(\Lambda^{-2}\right)={\bf y}}\atop
{{\bf r}(0)={\bf x}}}^{}{\cal D}{\bf r}
{\rm e}^{-\frac12\int\limits_{0}^{\Lambda^{-2}}d\tau\dot{\bf r}^2(\tau)}{\rm tr}{\,}\left[t^aU(r_{\bf xy})t^b
U(r_{\bf yx})\right],$$
where the parallel transportes are defined as in eq.~(\ref{u}), but along the regulator paths $r_{\bf xy}$ and
$r_{\bf yx}$, whose typical length is of the order $\Lambda^{-1}$. By virtue of this representation, we obtain the
following analogue of eq.~(\ref{2}):

$$\left(\gamma\partial_t+\Delta\right)W=$$

$$=\frac{g}{N}
\oint\limits_{C}^{} dx_i{\,}\int d^3y\int\limits_{{{\bf r}\left(\Lambda^{-2}\right)={\bf y}}\atop
{{\bf r}(0)={\bf x}}}^{}
{\cal D}{\bf r}
{\rm e}^{-\frac12\int\limits_{0}^{\Lambda^{-2}}d\tau\dot{\bf r}^2(\tau)}
(t^a)^{ij}(t^a)^{kl}(t^b)^{mn}\left<\zeta_i^b({\bf y},t)U^{lm}(r_{\bf xy})U^{nk}(r_{\bf yx})U_{\bf xx}^{ji}\right>.$$
Having in mind the eventual limit $\Lambda\to\infty$, we may use eq.~(\ref{deriv}) for the evaluation of the latter average.
Indeed, in this case, the use of the regularized expression instead of the $\delta$-function in eq.~(\ref{deriv}) would lead
to the excess of accuracy in the ($\Lambda\to\infty$)-limit (i.e, to the leading order in $1/\Lambda$ under study,
the resulting nonlocal expression will anyway be reduced
to the local one). By virtue of eq.~(\ref{tt}), we then obtain in the
large-$N$ limit:

$$\left(\gamma\partial_t+\Delta\right)W=\frac{g^2T}{N}
\oint\limits_{C}^{} dx_i{\,}\int d^3y\int\limits_{{{\bf r}\left(\Lambda^{-2}\right)={\bf y}}\atop
{{\bf r}(0)={\bf x}}}^{}
{\cal D}{\bf r}
{\rm e}^{-\frac12\int\limits_{0}^{\Lambda^{-2}}d\tau\dot{\bf r}^2(\tau)}
(t^b)^{mn}\left<\frac{\delta}{\delta A_i^b({\bf y},t)}U^{im}(r_{\bf xy})U^{nj}(r_{\bf yx})U_{\bf xx}^{ji}\right>.$$
In the evaluation of the last variational derivative, we may take
into account that the paths $r_{\bf xy}$ and $r_{\bf yx}$ are infinitesimal, i.e., to the leading order in
$1/\Lambda$ it is enough to differentiate only $U_{\bf xx}^{ji}$. Integrating then finally over $d^3y$ by virtue
of the resulting $\delta$-function, we arrive at the following regularized equation:

\begin{equation}
\label{regul}
\left(\gamma\partial_t+\Delta\right)W=\lambda T\oint\limits_{C}^{}dx_i\oint\limits_{C}^{}dy_i
\int\limits_{{{\bf r}\left(\Lambda^{-2}\right)={\bf y}}\atop
{{\bf r}(0)={\bf x}}}^{}
{\cal D}{\bf r}
{\rm e}^{-\frac12\int\limits_{0}^{\Lambda^{-2}}d\tau\dot{\bf r}^2(\tau)}
W[C_{\bf xy}r_{\bf yx},t]W[C_{\bf yx}r_{\bf xy},t].
\end{equation}
Clearly, in the limit $\Lambda\to\infty$, it recovers eq.~(\ref{3}) in the same way, as the regularized equation
derived in ref.~\cite{mh} recovers the usual large-$N$ QCD loop equation. Owing to the closeness of the contours
$C_{\bf xy}r_{\bf yx}$ and $C_{\bf yx}r_{\bf xy}$, eq.~(\ref{regul}) is formulated entirely on the loop space.

Let us now turn ourselves to the perturbative analysis of the formal solution to the Cauchy problem for eq.~(\ref{3}).
This solution obviously reads

\begin{equation}
\label{pert}
W\left[\vec x,t\right]=\frac{1}{\gamma}\int\limits_{0}^{t}d\tau{\rm e}^{\frac{\tau-t}{\gamma}\Delta}J\left[\vec x,\tau\right]+
{\rm e}^{-\frac{t}{\gamma}\Delta}W_0\left[\vec x{\,}\right],
\end{equation}
where $J\left[\vec x,\tau\right]$ is defined by the r.h.s. of eq.~(\ref{3}).
Let us assume that $\partial_tW$ vanishes at $t\to 0$ and justify this assumption later perturbatively.
Then in the limit $t\to 0$, eq.~(\ref{3}) takes the form of the usual 3D loop equation (with the only difference that the
dimensionful 3D 't Hooft coupling is replaced by $\lambda T$), which is obeyed by $W_0$. Owing to this fact, to the order $\lambda^1$,
$W_0$ can be represented as [cf. e.g. ref.~\cite{rev} and eq.~(\ref{D0})]~\footnote{Clearly, another assumptions imposed on the
limit $\lim_{t\to 0}^{}\partial_tW$ correspond to the choices of $W_0$ which
yield the perturbative expansions different from the one of eq.~(\ref{W0}).}

\begin{equation}
\label{W0}
W_0=1-\frac{\lambda T}{2}\oint\limits_{C}^{}dx_i\oint\limits_{C}^{}dy_iD_0({\bf x}-{\bf y})+{\cal O}\left(\lambda^2\right).
\end{equation}
Substituting further this expansion
together with the Ansatz $W=W^{(0)}+\lambda W^{(1)}$ into eq.~(\ref{pert}), we immediately find $W^{(0)}=1$,
while for $W^{(1)}$ we have

$$
W^{(1)}=\frac{Tt}{\gamma}\oint\limits_{C}^{}dx_i
\oint\limits_{C}^{}dy_i\delta({\bf x}-{\bf y})-\frac{T}{2}\left(1-\frac{t}{\gamma}\Delta\right)
\oint\limits_{C}^{}dx_i\oint\limits_{C}^{}dy_iD_0({\bf x}-{\bf y})=
-\frac{T}{2}\oint\limits_{C}^{}dx_i\oint\limits_{C}^{}dy_iD_0({\bf x}-{\bf y}).$$
In the derivation of this expression, apart from the formula
$\Delta\oint_{C}^{}dx_i
\oint_{C}^{}dy_i\delta({\bf x}-{\bf y})=0$, mentioned in the previous Section, we have again used
eq.~(\ref{D0}).
Thus, we see that $W=W_0$ up to the first order of the expansion in $\lambda$. In particular,
this fact justifies our initial assumption $\lim_{t\to 0}^{}\partial_tW=0$ at least up to this order of $\lambda$.
The $t$-dependence of $W$ appears in the next orders of perturbation theory. This is because the action of $\Delta^n$,
($n=2,\ldots,n_{\rm max}$, where $n_{\max}$ is determined by the order of perturbation theory)
onto the respective terms of the expansion of $W_0$ gives then a nonvanishing result.

\section{Summary}
In the present letter, we have derived and analysed the large-$N$ loop equation in quark-gluon plasma. This
equation is based on the Langevin equation which describes the dynamics of soft degrees of freedom. The regularized
form of the loop equation, formulated  entirely on the loop space, has also been obtained. It is derivable directly
from the regularized version of the original Langevin equation. By making use of the method of smearing of the loop-space Laplacian,
we have further solved the Cauchy problem for the unregularized equation in some particular case. Namely, this has been done for the
gauge potential of the form $A_i={\bf n}{\bf H}\psi_i$, where ${\bf H}$'s are the Cartan generators of the $SU(N)$-group,
${\bf n}$ is a $(N-1)$-dimensional constant unit vector, and $\psi_i({\bf x}, t)$ is an arbitrary 3D vector-function.
The obtained solution is approximate and valid up to terms of the order
${\cal O}\left(\left(\frac{\tau}{\gamma S_{\rm min}}\right)^2\right)$, which are small for the characteristic temporal and spatial scales of
soft fields. We have also briefly addressed the perturbative expansion of the formal solution to the Cauchy problem
in the general case (i.e., for an arbitrary gauge potential). Clearly, to perform this analysis, one first needs to fix
the perturbative expansion for the initial value of the Wilson loop. In order to extract such an expansion from the
obtained loop equation, it is necessary to postulate a certain spatial dependence for $\lim_{t\to 0}^{}\partial_tW$. In the simplest case,
when this limit is set to zero, the loop equation for the initial value of the Wilson loop takes the functional
form of the usual large-$N$ QCD loop equation. The latter is known~\cite{le, rev} to be satisfied by the
perturbative expansion of the Wilson loop, computed in large-$N$ QCD. Using this expansion, it has been checked that
for the case  $\lim_{t\to 0}^{}\partial_tW=0$ under study and to the first order in the
't~Hooft coupling, the solution to the Cauchy problem is equal to the initial value of the Wilson loop at any times.
(In particular, this fact justifies the postulate $\lim_{t\to 0}^{}\partial_tW=0$ itself, within this order of perturbation theory.)
However, even in such a case, the time-dependence of the solution does appear in the higher orders of perturbation theory.

\section*{Acknowledgments}
The author is grateful to Profs.
A.~Di~Giacomo and Yu.M.~Makeenko for
useful discussions. He is also grateful to Prof. A.~Di~Giacomo and to the
whole Theory Group of the Physics Department of the Pisa University
for cordial hospitality. This work has been supported by INFN and partially by the
INTAS grant Open Call 2000, project No. 110.



\begin{thebibliography}{99}
\bibitem{le}
Yu.M. Makeenko and A.A. Migdal, Phys. Lett. {\bf B 88} (1979) 135;
Nucl. Phys. {\bf B 188} (1981) 269;
Yu.M. Makeenko, Sov. J. Nucl. Phys. {\bf 33} (1981) 274.

\bibitem{rev}
A.A. Migdal, Phys. Rep. {\bf 102} (1983) 199.

\bibitem{str}
A.A. Migdal, Nucl. Phys. {\bf B 189} (1981) 253;
A.M. Polyakov, Nucl. Phys. {\bf B 486} (1997) 23;
A.M. Polyakov and V.S. Rychkov, Nucl. Phys. {\bf B 581} (2000) 116; ibid. {\bf B 594} (2001) 272.


\bibitem{dgo}
N. Drukker, D.J. Gross, and H. Ooguri, Phys. Rev. {\bf D 60} (1999) 125006.

\bibitem{susy}
Yu.M. Makeenko and P.B. Medvedev, Nucl. Phys. {\bf B 193} (1981) 444.

\bibitem{gravity}
Yu.M. Makeenko and N.A. Voronov, Sov. J. Nucl. Phys. {\bf 36} (1982) 444.

\bibitem{tur1}
A.A. Migdal, Int. J. Mod. Phys. {\bf A 9} (1994) 1197.

\bibitem{tur2}
D.V. Antonov, Mod. Phys. Lett. {\bf A 11} (1996) 3113.

\bibitem{bod}
D. B\"odeker, Phys. Lett. {\bf B 426} (1998) 351.

\bibitem{sg}
A. Selikhov and M. Gyulassy, Phys. Lett. {\bf B 316} (1993) 373.

\bibitem{asy}
P. Arnold, D.T. Son, and L.G. Yaffe, Phys. Rev. {\bf D 59} (1999) 105020.

\bibitem{zj}
J. Zinn-Justin,
{\it Quantum field theory and critical phenomena} (2nd edn., Oxford Univ. Press,
New York, 1993).

\bibitem{group}
R. Gilmore, {\it Lie groups, Lie algebras, and some of their applications}
(J. Wiley \& Sons, New York, 1974).

\bibitem{lapl}
Yu.M. Makeenko, Phys. Lett. {\bf B 212} (1988) 221; preprints ITEP 88-50, ITEP 18-89,
{\it Large-N QCD on Loop Space} (unpublished).

\bibitem{bhst}
Z. Bern, M.B. Halpern, L. Sadun, and C. Taubes, Nucl. Phys. {\bf B 284} (1987) 1, 35.

\bibitem{mh}
M.B. Halpern and Yu.M. Makeenko, Phys. Lett. {\bf B 218} (1989) 230.

\end{thebibliography}
\end{document}